\documentclass[%
 reprint,
superscriptaddress,
 amsmath,amssymb,
 aps,
prb,
]{revtex4-2}

\usepackage[disable]{todonotes} 

\usepackage{graphicx}
\usepackage{dcolumn}
\usepackage{bm}

\begin{document}

\preprint{APS/123-QED}

\title{Effects of Strain Compensation on Electron Mobilities in InAs Quantum Wells\\Grown on InP(001)}

\author{C.P. Dempsey}
\email{c\_dempsey@ucsb.edu}
\affiliation{%
 Department of Electrical and Computer Engineering,  University of California, Santa Barbara, CA 93106
}%
\author{J.T. Dong}%
\affiliation{%
 Materials Department, University of California\\Santa Barbara, Santa Barbara, CA 93106, USA
}%

\author{I. Villar Rodriguez}%
\affiliation{%
 London Centre for Nanotechnology, University College London, 17-19 Gordon Street, London WC1H 0AH, United Kingdom
}%

\author{Y. Gul}%
\affiliation{%
 London Centre for Nanotechnology, University College London, 17-19 Gordon Street, London WC1H 0AH, United Kingdom
}%

\author{S. Chatterjee}
\affiliation{%
 Department of Electrical and Computer Engineering,  University of California, Santa Barbara, CA 93106
}%

\author{\\M. Pendharkar}
\altaffiliation[Current address: ]{Department of Materials Science and\\Engineering, Stanford University, Stanford, CA 94305
}%
\affiliation{%
 Department of Electrical and Computer Engineering,  University of California, Santa Barbara, CA 93106
}%

\author{S.N. Holmes}%
\affiliation{%
 Department of Electronic and Electrical Engineering, University College London, Torrington Place, London WC1E 7JE, United Kingdom
}%

\author{M. Pepper}%
\affiliation{%
 London Centre for Nanotechnology, University College London, 17-19 Gordon Street, London WC1H 0AH, United Kingdom
}%
\affiliation{%
 Department of Electronic and Electrical Engineering, University College London, Torrington Place, London WC1E 7JE, United Kingdom
}%

\author{C.J. Palmstr\o m}
\email{cjpalm@ucsb.edu}
\affiliation{%
 Department of Electrical and Computer Engineering,  University of California, Santa Barbara, CA 93106
}%
\affiliation{Materials Department, University of California\\Santa Barbara, Santa Barbara, CA 93106, USA
}%

\date{\today}

\begin{abstract}
InAs quantum wells (QWs) grown on InP substrates are interesting for their applications in devices with high spin-orbit coupling (SOC) and their potential role in creating topologically nontrivial hybrid heterostructures. These QWs rely on InGaAs cladding layers and InAlAs barrier layers to confine electrons within a thin InAs well. The highest mobility QWs are limited by interfacial roughness scattering and alloy disorder scattering in the cladding and buffer layers. Increasing QW thickness has been shown to reduce the effect of both of these scattering mechanisms. However, for current state-of-the-art devices with As-based cladding and barrier layers, the critical thickness is limited to $\leq7$ nm. In this report, we demonstrate the use of strain compensation techniques in the In$_x$Ga$_{1-x}$As cladding layers, grown on In$_{0.81}$Al$_{0.19}$As barrier layers, to extend the critical thickness well beyond this limit. We induce tensile strain in the InGaAs cladding layers by reducing the In concentration from In$_{0.81}$Ga$_{0.19}$As to In$_{0.70}$Ga$_{0.30}$As and we observe changes in both the critical thickness of the well and the maximum achievable mobility. The peak electron mobility at 2 K is $1.16\times10^6$ cm$^2/$Vs, with a carrier density of $4.2\times10^{11}$ /cm$^2$. Additionally, we study the quantum lifetime and Rashba spin splitting in the highest mobility device as these parameters are critical to determine if these structures can be used in topologically nontrivial devices.
\end{abstract}

\maketitle


\section{\label{sec:Introduction}Introduction}
Bulk InAs has a Land\'e g-factor of 15 \cite{Pidgeon1967}, making it a useful material for studying large spin-orbit coupling in devices. Additionally, QWs of InAs have high mobilities ($>1\times10^6$ cm$^2$/Vs) \cite{Hatke2017,Thomas2018,Tschirky2017}, and near surface QWs have demonstrated gateable induced superconductivity \cite{Shabani2016}, making them an ideal platform for the study of topologically nontrivial hybrid heterostructures. InAs quantum wells have been grown on GaAs, InP, and nearly lattice-matched GaSb substrates. A record mobility of $2.4\times10^6$ cm$^2$/Vs was achieved in a 21 nm thick InAs QW utilizing Sb-based barriers (Al$_{x}$Ga$_{1-x}$Sb and/or AlSb barrier layers \cite{Tschirky2017,Thomas2018}), a GaSb substrate, and by tuning the carrier density to n$_s\sim1\times10^{12}$ \cite{Tschirky2017}. Shojaei \textit{et al.} \cite{Shojaei2016} demonstrated that interface roughness was one of the dominant scattering mechanisms at high carrier densities and that these effects could be mitigated by increasing the QW width, leading to an increase in the maximum mobility. Sb-based InAs QW structures are utilized because the cladding layers, barrier layers, and substrate are all nearly lattice matched with the InAs layer, increasing the critical thickness of the QW. However, when Hall bar devices are fabricated, InAs QWs with Sb-based cladding and barrier layers typically suffer from side wall conduction \cite{Mueller2017}. Different methods of passivation have been experimented with, but the trivial edge conductance could not be completely removed from the transport spectra \cite{Mittag2017}.

Trivial edge conductance is not observed for Hall bars of InAs QWs with As-based cladding and barrier layers. However, these structures are typically grown on an InP substrate. InAs grown pseudomorphically on InP is under a compressive strain of $-3.1\%$. This large lattice mismatch requires the use of metamorphic buffer layers to increase the in-plane lattice constant of the film surface while minimizing the nucleation of threading dislocations. Buffer layer engineering has enabled these QWs to reach a record mobility of $\mu=1.1\times10^6$ cm$^2/$Vs at a carrier density of n$_s=6.2\times10^{11}$ /cm$^2$ \cite{Hatke2017}, minimizing the difference between InAs QWs grown on InP substrates and InAs QWs grown on GaSb substrates. However, these current state-of-the-art InAs QWs grown on InP are limited to thicknesses of 4 nm due to the in-plane lattice constant difference between the InAs layer and the surrounding In$_{x}$Ga$_{1-x}$As cladding and In$_{x}$Al$_{1-x}$As barrier layers. When more than $4$ nm of InAs is grown on an $x=0.75$ In-containing active region, the QW begins to relax, lowering the sample mobility \cite{Hatke2017}. Previous work \cite{Shabani2016,Benali2022} has demonstrated that by increasing the In concentration of the In$_{x}$Ga$_{1-x}$As cladding/In$_{x}$Al$_{1-x}$As barrier layers from $x=0.75$ to $x=0.81$, the critical thickness of the InAs layer can be increased up to 7 nm. Additionally, Benali \textit{et al.} \cite{Benali2022} demonstrated this increase leads to a higher fraction of the 2D electron gas (2DEG) being confined within the InAs layer, lowering its penetration into the cladding and barrier layers, and reducing the effects of alloy disorder and interfacial roughness scattering. In doing so, they proceeded to break the record for the highest mobility of InAs QWs grown on a GaAs substrate \cite{Benali2022}. While further increases in the In content of the barrier and cladding layers would enable thicker InAs QWs, it would also decrease QW confinement, potentially leading to an increase in scattering and minimizing any potential benefits.

Strain compensation provides another pathway to increase QW critical thickness beyond the current technical limits of 7 nm without sacrificing 2DEG confinement. Early uses of strain compensation include increasing the maximum number of QWs in multi quantum well structures \cite{Seltzer1991,Tsuchiya1994,Tu1995} by growing InGaAsP layers under tensile strain to counteract the compressive strain of the In$_{0.67}$Ga$_{0.33}$As QW layers. This produced a zero net strain structure and demonstrated the promise of strain compensation in increasing critical thickness. In this paper we discuss the effects of implementing strain compensation by tuning the In concentration of the cladding layers (In$_{x}$Ga$_{1-x}$As layers) and the impact that strain compensation has on both the maximum thickness of the QW and the maximum mobility of the heterostructure.

\begin{figure*}[t]
    \includegraphics[width=7in,clip,keepaspectratio]{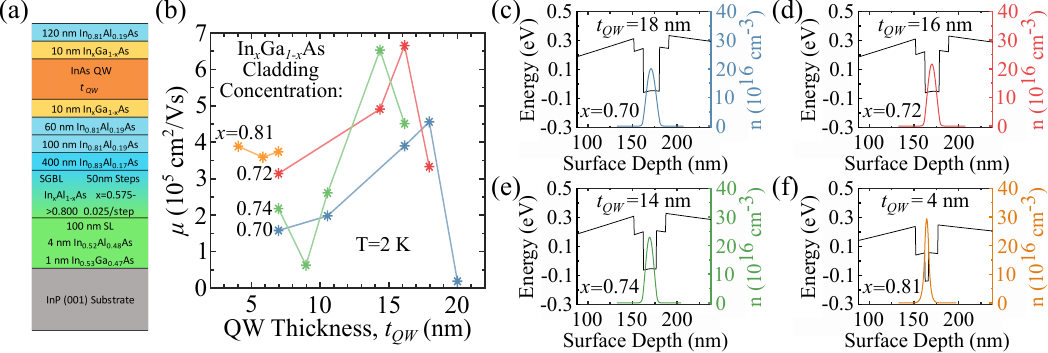}
	\caption{(a) Schematic representation of InAs QW layer structure. (b) van der Pauw mobilities for four series of samples with differing strain levels in the In$_x$Ga$_{1-x}$As cladding layers and QW thicknesses ($t_{QW}$). The corresponding carrier density for each sample is depicted in Fig. S2 of the Supporting Information \cite{supp}. (c)-(f) 1D Schr\"odinger-Poisson simulations \cite{Tan1990} calculated for the highest mobility QW structure, (see Table \ref{tab:samples} for detailed properties) from each series of In$_x$Ga$_{1-x}$As. As QW thickness decreases, the percentage of carriers confined in the InAs layer is reduced.}
    \label{fig:qwIntro}
\end{figure*}
 \section{\label{sec:Methods}Methods}
The QW samples studied in this paper were grown by molecular beam epitaxy (MBE) in a VG V80H growth chamber. Fe-doped InP $(001)$ semi-insulating epi-ready substrates were purchased from AXT Inc. The native oxide on the InP substrate was thermally desorbed by heating the substrate under an As$_2$ flux, generated by a valved cracker source, while the reconstruction was monitored using reflection high energy electron diffraction (RHEED). The transition to the metal rich $(4\times2)$ reconstruction was used to calibrate the pyrometer temperature to 515 $^\circ$C and upon observing the transition, the sample temperature was immediately reduced to 480 $^\circ$C for the nucleation of a lattice-matched In$_{0.52}$Al$_{0.48}$As/In$_{0.53}$Ga$_{0.47}$As superlattice. After growth of the superlattice, the sample temperature was further reduced to 340 $^\circ$C to grow the step graded buffer layers (SGBL). Each layer of the SGBL is 50 nm thick and the In concentration is increased by 2.5$\%$ for each  step, up to In$_{0.80}$Al$_{0.20}$As. Next, a 400 nm overstep layer of In$_{0.83}$Al$_{0.17}$As was grown to fully relieve residual strain in the SGBL. Finally, a 100 nm layer of In$_{0.81}$Al$_{0.19}$As was grown, utilizing the same composition as Benali \textit{et al.} \cite{Benali2022} for the active region. At this point, growth of the active region of the QW commenced and the substrate was heated to 450 $^\circ$C.

A 60 nm thick In$_{0.81}$Al$_{0.19}$As bottom barrier layer was grown with a 10 nm thick In$_{x}$Ga$_{1-x}$As bottom cladding layer grown subsequently. The strain in the cladding layers was modified by changing $x$ from $x=0.81$ to $0.74$, $0.72$, or $0.70$. By reducing $x$, the lattice constant of the In$_{x}$Ga$_{1-x}$As decreases relative to the In$_{0.81}$Al$_{0.19}$As bottom barrier, leading to an accumulation of tensile strain energy within the cladding layer. Next, the InAs QW, which will always be under compressive strain in these structures, was grown with thickness, $t_{QW}$. By varying the lattice constant of the cladding layers, $t_{QW}$ was increased beyond the critical thicknesses, $h_c$, previously found for InAs QWs grown with As-based barriers ($h_c\sim4$ nm for In$_{0.75}$Al$_{0.25}$As \cite{Hatke2017} and $h_c\sim7$ nm for In$_{0.81}$Al$_{0.19}$As \cite{Benali2022}). Finally, a top cladding layer of 10 nm In$_{x}$Ga$_{1-x}$As and a 120 nm thick In$_{0.81}$Al$_{0.19}$As top barrier layer were grown to complete the QW structure, as depicted in Fig. \ref{fig:qwIntro}(a).

Low temperature electron transport measurements were carried out utilizing a Quantum Design Physical Property Measurement System equipped with a magnet capable of generating fields up to 14 T. Sample resistance was determined using low frequency AC lock-in measurements with an AC excitation of $1\ \mu$A. Samples were measured using the van der Pauw (vdP) configuration to determine which samples had the highest mobility based on the In concentration in the cladding layer and the QW thickness, $t_{QW}$. These samples were created by cleaving the QWs into approximately square pieces prior to soldering In-Sn contacts onto the corners and annealing for 15 minutes at 250 $^\circ$C. The highest mobility samples were fabricated into Hall bars with the long axis oriented along the $[1\bar{1}0]$. The mesa was formed using an etchant of H$_2$SO$_4$:H$_2$O$_2$:H$_2$O (1:8:120). Next, NiAuGe contacts were deposited and annealed at 450 $^\circ$C for 120s under forming gas to create ohmic contacts between the NiAuGe and the quantum well. A 30 nm Al$_2$O$_3$ gate dielectric was deposited using atomic layer deposition. Vias were defined to the NiAuGe contact pads, by etching through the dielectric. Finally, a layer of Ti/Au was deposited to fill the vias and form the top gate, enabling control of the carrier concentration within the QW. The fabricated Hall bar has an arm spacing of 740 $\mu$m and a width of 80 $\mu$m. These samples were used for gated magnetotransport measurements and characterization of the thermal dependence of the Shubnikov de Haas (SdH) oscillations. Additionally, Rashba measurements were conducted utilizing a magnetic field modulation technique with two separate solenoids. The larger solenoid provides a DC magnetic field while the smaller solenoid provides a small AC field, enabling the direct measurement of $dR_{xx}/dB$ and $d^2R_{xx}/dB^2$.

\begin{table}[b]
  \begin{tabular}{| c | c | c | c |}
    \hline
    Sample & $x$ Conc. of & QW Thickness & Cladding \\
     &In$_x$Ga$_{1-x}$As & $t_{QW}$ (nm) & Misfit (\%) \\ \hline \hline
    Sample A & $0.70$ & $18$ & 0.78\\ \hline
    Sample B & $0.72$ & $16$ & 0.64\\ \hline
    Sample C & $0.74$ & $14$ & 0.50\\ \hline
    Sample D & $0.81$ & $4$ & 0.03 \\ \hline
  \end{tabular}
  \caption{Highest quality samples from each In concentration series depicted in Fig. \ref{fig:qwIntro}(b) with relevant properties listed. Cladding misfit is calculated using the formula:\\ $f=(a_{barrier}-a_{cladding})/a_{cladding}$}\label{tab:samples}
\end{table}

\section{\label{sec:Results}Results and discussion}
\subsection{\label{sec:strucSim}Tuning Cladding Layer Properties}
Magnetotransport measurements were performed on four series of samples with In$_{0.81}$Al$_{0.19}$As barrier layers and In$_x$Ga$_{1-x}$As cladding layers. The In concentration ($x$) of the cladding layer was varied for each series with $x$ set to $0.70,\ 0.72,\ 0.74,$ or $0.81$, changing the tensile strain present in the cladding layer. Additionally, the quantum well thickness, $t_{QW}$, was varied to determine whether strain compensation could enable the growth of thicker InAs QWs. These samples were wired in the vdP geometry, and magnetic field sweeps were utilized to extract the relationship between the In concentration in the cladding layer, the maximum thickness of the QW, and the mobility of the QW, as depicted in Fig. \ref{fig:qwIntro}(b). The samples included InAs QWs grown on In$_{0.81}$Ga$_{0.19}$As (which has a misfit of only $0.03\%$, as defined in Table \ref{tab:samples}) cladding layers with thicknesses ranging from $t_{QW}=4-7$ nm. These samples enable us to compare our QWs with other state-of-the-art QWs as previous growths of InAs on $x=0.75$ In-containing active regions were limited to a thickness of 4 nm \cite{Hatke2017} and growth of InAs layers on $x=0.81$ In-containing active regions were limited to a thickness of 7 nm \cite{Benali2022}. We observe minimal change in mobility as we increase $t_{QW}$ for the $x=0.81$ series. However, this is partially due to lack of control over the carrier concentration present in the QW and slight differences in thermal desorption temperatures, which can translate to variance in the QW mobility. Gated Hall bar measurements (not shown) demonstrated that the peak gated mobility of the sample with $t_{QW}=7$ nm is greater than that of the $t_{QW}=4$ nm sample, as expected from the results of Benali \cite{Benali2022}.

After studying the samples grown on In$_{0.81}$Ga$_{0.19}$As cladding layers, the In content in the cladding layers was reduced. All of the 7 nm thick QWs grown on strained cladding layers had lower mobilities relative to the samples grown on In$_{0.81}$Ga$_{0.19}$As. This may result from the increased alloy disorder scattering present as the In concentration, $x$, of the cladding layers is reduced. Previous work on InGaAs QWs has demonstrated that alloy disorder scattering plays an important role in the maximum achievable sample mobility \cite{Capotondi2005_2,Dong2024}.

As the QW thickness is increased beyond 7 nm, the measured vdP mobility generally increases in all of the samples until a peak mobility is achieved. To better understand this relationship, we utilized a 1D Schr\" odinger-Poisson simulation \cite{Tan1990} to calculate the band structure and carrier concentrations for the highest mobility sample from each In concentration series measured in Fig. \ref{fig:qwIntro}(b) (The structural properties of these selected samples are summarized in Table \ref{tab:samples}). The results can be seen in Figs. \ref{fig:qwIntro}(c)-(f). By increasing QW thickness, the percentage of 2DEG contained within the InAs layer increases. The $18$ nm InAs layer contains $97.2\%$ of the charge carriers, the $16$ nm well contains $96.8\%$, and the $14$ nm well contains $95.0\%$. On the other hand, for the $4$ nm thick QW, the percentage of the carrier density residing within the InAs layer drops to $50.2\%$. These simulations demonstrate that 2DEG confinement has a strong dependence on the width of the InAs layer. Additionally, these simulations show that by increasing QW thickness, the effects of interfacial roughness and alloy disorder scattering may be reduced as the carriers in thinner wells, like the 4 nm QW, reside much closer to the InGaAs/InAlAs regions. This explains the relationship between increasing $t_{QW}$ and the measured QW mobilities in Fig. \ref{fig:qwIntro}(b).

After the peak mobility is achieved, a sharp drop off in mobility is observed. It is interesting to note the peak mobility position shifts to thicker QWs as the cladding layer lattice misfit is increased (misfit is defined as $f=(a_{barrier}-a_{cladding})/a_{cladding}$, where $a_{cladding}$ is the cladding layer lattice constant and $a_{barrier}$ is the barrier layer lattice constant), indicating a correlation between the misfit of the cladding layer and the peak mobility. This relationship suggests that the drop in mobility is likely caused by relaxation of the InAs QW, similar to the growth of a 6 nm QW on an $x=0.75$ In-containing active region grown by Hatke \textit{et al.} \cite{Hatke2017}. Additionally, this relationship demonstrates that the In$_{x}$Ga$_{1-x}$As cladding layers are effectively offsetting the lattice constant mismatch between the InAs layer and the In$_{0.81}$Al$_{0.19}$As barrier layers, delaying the onset of QW relaxation, as intended for strain compensated structures \cite{Seltzer1991,Tu1995,Tsuchiya1994}. The peak mobility found for the In$_{0.70}$Ga$_{0.30}$As cladding layers is much lower than the peak mobility determined for either the In$_{0.72}$Ga$_{0.28}$As or In$_{0.74}$Ga$_{0.26}$As series of QWs. We interpret this to be a result of lattice relaxation and dislocation formation as reported by Hatke \textit{et al.} \cite{Hatke2017} caused by the larger misfit of the In$_{0.70}$Ga$_{0.30}$As cladding layers.

\subsection{\label{sec:gatedT}Gated Transport Measurements}
\begin{figure}[t]
    \includegraphics[width=3.4in,clip,keepaspectratio]{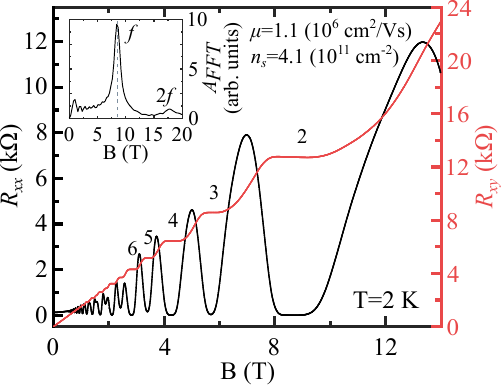}
	\caption{Magnetotransport results of Sample B. Clear oscillations are present in $R_{xx}$ and low field $R_{xy}$ is linear indicating single channel transport. Inset: FFT of background-removed oscillations from low field $R_{xx}$ measurements. A single peak and its harmonic are detected confirming single channel transport.}
    \label{fig:qwOsc}
\end{figure}

The longitudinal and Hall resistance of the highest mobility sample, Sample B (see characteristics in Table \ref{tab:samples}), are depicted in Fig. \ref{fig:qwOsc}. Voltage was applied to the top gate to set the carrier concentration to $4.1\times10^{11}\ /$cm$^2$. At this carrier concentration, the sample mobility was measured to be $\mu=1.14\times10^{6}\ $cm$^2/$Vs. At low fields, the Hall effect is found to be linear and at high fields, the longitudinal resistance reaches 0 while the Hall resistance plateaus at quantized resistance values, indicating that no parallel transport channels are present in the QW structure. We confirm single channel transport utilizing SdH oscillations, which are well-developed in the low field $R_{xx}$ trace. Carrier concentration is extracted from the SdH oscillations by removing the background of $R_{xx}$ using a linear fit. The resulting oscillations are Fourier transformed and the result is depicted in the inset of Fig. \ref{fig:qwOsc}. The single peak and its harmonic in the FFT indicate that one transport channel is contributing to the SdH oscillations observed. We extracted the frequency of the inverse field oscillations and used the Onsager relation \cite{Onsager1952} to relate the frequency of the oscillations to the 2D carrier density of the transport channel. We found $n_{SDH}=4.06\times10^{11}\ /$cm$^2$, which agrees with the carrier concentration extracted from the Hall measurement. 

\begin{figure}[t]
    \includegraphics[width=3.4in,clip,keepaspectratio]{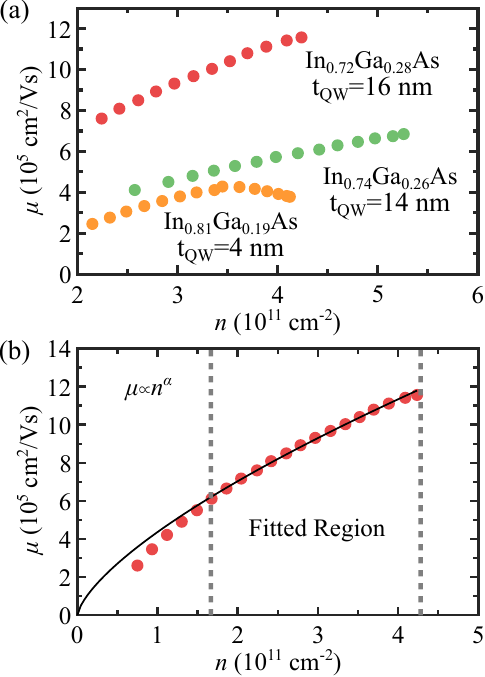}
	\caption{(a) Mobility vs. carrier density for the highest mobility samples from the $x=0.72$, $0.74$, and $0.81$ In$_x$Ga$_{1-x}$As series measured in Fig. \ref{fig:qwIntro}(b). (b) Extended mobility vs. carrier density for Sample B, the highest mobility QW measured. The black line is a power-law fit using the expression: $\mu\propto n_s^\alpha$.}
    \label{fig:qwHBComp}
\end{figure}

Magnetotransport measurements were taken at a wide range of gate voltages for the highest mobility sample from each of the $x=0.72$, $0.74$, and $0.81$ In$_x$Ga$_{x-1}$As series and the results are depicted in Fig. \ref{fig:qwHBComp}(a). By applying voltage to the top gate, we are able to modulate the carrier density within the QWs. The enhancement of mobility with increasing carrier density is due to better screening of charged defects, reducing the effects of Coulomb scattering \cite{Bastard1988}. Sample B (a 16 nm QW) shows the highest mobility of all of the samples studied and by increasing the carrier concentration within the well we are able to increase the mobility to a maximum of $\mu=1.16\times10^6$ cm$^2/$Vs, which is believed to be the highest mobility recorded for any InAs QW grown on an InP substrate that we can find in the literature \cite{Hatke2017}. The maximum mobility occurs at a carrier density of $n_s=4.2\times10^{11}$ /cm$^2$. This represents a nearly 20$\%$ increase in mobility when compared to the record breaking QW grown by Hatke \textit{et al.} \cite{Hatke2017} for a similar carrier density. The sample with the next highest mobility is Sample C (a 14 nm QW), with a maximum recorded mobility of $\mu=6.85\times10^5$ cm$^2/$Vs. The difference between the maximum attainable mobilities of sample B and sample C likely relates to the onset of relaxation. As shown by Hatke et al. the increase of the InAs layer from 4 nm to 6 nm led to a halving of the mobility \cite{Hatke2017}. The thickness of the QW in sample C may have exceeded the critical thickness of InAs by 3-4 nm, leading to subsequent relaxation and a large reduction in QW mobility. The relationship of the carrier density to the mobility beyond the peak for both the In$_{0.72}$Ga$_{0.28}$As QW and the In$_{0.74}$Ga$_{0.26}$As QW is not depicted in Fig. \ref{fig:qwHBComp}(a), but is further discussed in the Supporting Information, Fig. S4 \cite{supp}.

To further understand the limits to the carrier mobility in Sample B, we studied the relationship between carrier density and mobility over a larger number of carrier concentrations as depicted in Fig. \ref{fig:qwHBComp}(b). By tuning the top gate voltage between $-0.9$ V and $+0.1$ V, we are able to modulate the carrier density from $7.5\times10^{10}\ /$cm$^2$ up to $4.2\times10^{11}\ /$cm$^2$. In high-mobility two-dimensional systems the carrier mobility typically shows a power-law dependent relationship with the carrier density \cite{DasSarma2013}, $\mu\propto n_s^\alpha$. For the case of 2D carriers with 3D impurities and strong screening $(q_{TF}>>2k_F$, where $q_{TF}$ is the Thomas-Fermi wave vector and $k_F$ is the Fermi wave vector), $\alpha=0.5$ and for the case of 2D carriers with 3D impurities and weak screening $(q_{TF}<<2k_F)$, $\alpha=1.5$. Using a log-log plot for the relationship between $\mu$ and $n_s$, we fit the slope of the data between $n_s=1.6\times10^{11}$ and $n_s=4.2\times10^{11}$. We extract $\alpha=0.69$, which lies in between the two expected values of $\alpha$. This intermediate value of $\alpha$ suggests that the mobility in this regime is limited by scattering from nearby background charged impurities. Similar values have been observed previously in both InAs QWs grown on InP substrates \cite{Shabani2014} and InAs QWs grown on GaSb substrates \cite{Thomas2018}.

\subsection{\label{sec:qLife}Extracting Quantum Lifetime}

\begin{figure}[t]
    \includegraphics[width=3.15in,clip,keepaspectratio]{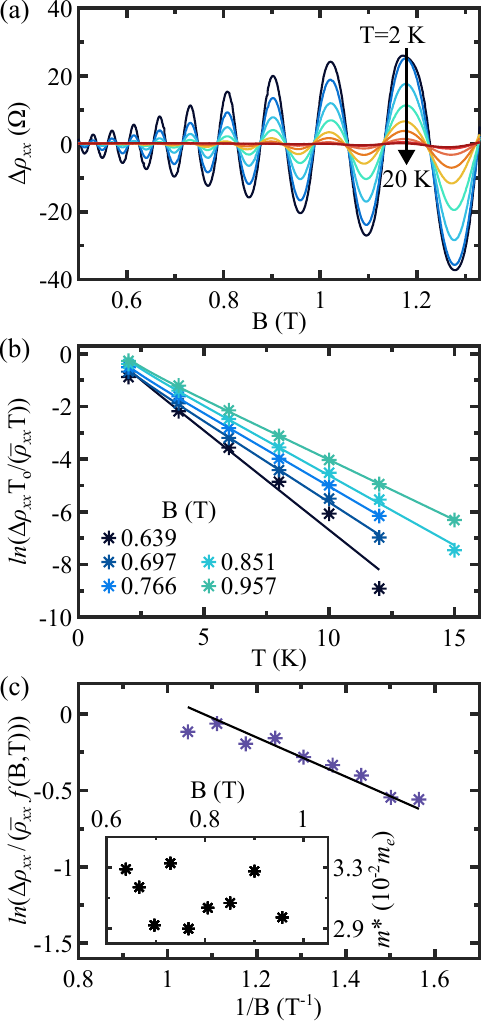}
	\caption{(a) Thermal damping of SdH oscillations in $\rho_{xx}$ after background has been subtracted by a linear fit. (b) $ln(\Delta\rho_{xx}T_o/(\bar{\rho}_{xx}T))$ vs. T for selected field amplitudes. Lines correspond to fits of data with $-ln(\sinh(2\pi^2k_BTm^*/\hbar q B))$, enabling extraction of $m^*$. (c) $ln(\Delta\rho_{xx}/(\bar{\rho}_{xx}f(B,T)))$ vs. $1/B$ at 2 K. The slope of the fitted black line is inversely proportional to the quantum lifetime. Inset: Extracted $m^*$ values for varying B field positions from (b).}
    \label{fig:effMass}
\end{figure}

Quantum lifetime has been suggested as a predictor of the strength of fractional quantum Hall states and provides insight into the prevalent scattering mechanisms affecting transport \cite{Qian2017}. To extract the quantum transport properties of Sample B (the 16 nm QW), we measured the longitudinal resistivity, $\rho_{xx}$, at a series of low temperatures ranging from $2$ K to $20$ K. A smooth polynomial was used to remove the resulting background. The resulting traces are depicted in Fig. \ref{fig:effMass}(a). The oscillation amplitude decreases as the temperature is increased as a result of larger energy averaging occurring around the Fermi energy and can be described by the resistivity function derived by Isihara and Smr\v cka \cite{Isihara1986,Laikhtman1994,Ihn2010}:
\begin{equation}
    \frac{\rho_{xx}}{\rho_o}=1+4e^{-\pi/\omega_c\tau_q}\frac{X}{\sinh(X)}\cos\bigg(2\pi\frac{hn}{2qB}+\pi\bigg).
    \label{eq:rhoxx}
\end{equation}
Where $\rho_o$ describes the zero field resistivity, $\tau_q$ is the quantum lifetime, $\omega_c$ is the cyclotron mass, defined as $\omega_c=qB/m^*$, X is the thermal dependence term, defined as $X=2\pi^2k_BT/\hbar\omega_c$, T is the temperature, n is the carrier concentration, and $B=\mu_o$H, where $\mu_o$ is the vacuum permeability, and H is the applied field. The only temperature dependent term in Eq. \ref{eq:rhoxx} is the $X/\sinh(X)$ term.
This allows us to extract the effective mass by plotting $ln(\Delta\rho_{xx}T_o/(\bar{\rho}_{xx}T))$ against T, as depicted in Fig. \ref{fig:effMass}(b) for specific field values \cite{Lei2020}. In this equation, $\Delta\rho_{xx}$ is the amplitude of the oscillation, $\bar{\rho}$ is the magnitude of the non-oscillatory background, and $T_o$ is the temperature at which the lowest field sweep was conducted \cite{Lei2020}. We extract the oscillation amplitude of multiple valleys and peaks before the onset of spin-splitting and selected results are depicted across multiple different temperatures in Fig. \ref{fig:effMass}(b). The effective mass is found by fitting the data with $-ln(\sinh(2\pi^2k_BTm^*/\hbar q B))$ and an offset \cite{Lei2023}. The results of the fit are also displayed in \ref{fig:effMass}(b). We determine $m^*=(0.031\pm0.002)m_e$, which agrees well with values found in the literature \cite{Shabani2014}. The inset of Fig. \ref{fig:effMass}(c) depicts the extracted effective mass values for different magnetic field intensities.

Utilizing the extracted effective mass we can now determine the quantum lifetime term, $\tau_q$, by plotting $ln(\Delta\rho_{xx}/(\bar{\rho}_{xx}f(B,T)))$ against the inverse field, where $f(B,T)=2\pi^2k_BT/\hbar\omega_c/\sinh(2\pi^2k_BT/\hbar\omega_c)$. The results are depicted in Fig. \ref{fig:effMass}(c). A single parameter linear fit can be applied to the data. The y-intercept is set to equal $ln(4)$ based on work by Coleridge \cite{Coleridge1991} and the slope of the line corresponds to: slope$=-\pi m^*/q\tau_q$ \cite{Thomas2018}. We find $\tau_q=0.45$ ps, which gives us a Dingle ratio of $\tau_t/\tau_q=45$, where $\tau_t=\mu m^*/q$ is defined as the scattering time. Similar Dingle ratios have been previously observed in InAs QWs grown on GaSb substrates \cite{Tschirky2017} with the same carrier concentration. The large difference in time constants is not unexpected as $\tau_t$ is primarily impacted by large angle scattering events. $\tau_q$, on the other hand, is affected equally by all scattering events \cite{Peters2016,Gold1988}.

\begin{figure}[t]
    \includegraphics[width=3.24in,clip,keepaspectratio]{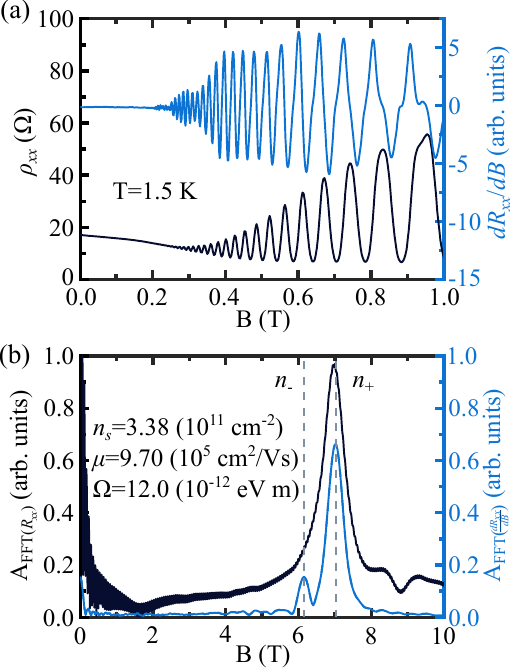}
	\caption{(a) \textit{Left axis:} Response of $\rho_{xx}$ to an applied magnetic field. \textit{Right axis:} Modulated magnetic field measurement of $dR_{xx}/dB$ showing beating patterns indicative of spin splitting. (b) \textit{Left axis:} FFT of the $\rho_{xx}$ data shown in (a) with a single peak evident. \textit{Right axis:} FFT of the $dR_{xx}/dB$ depicting two spin split peaks, labeled $n_-$ and $n_+$. Vertical dashed gray lines have been added as guides to the eye.}
    \label{fig:rashba}
\end{figure}
\subsection{\label{sec:SoCP}Determining Spin-Orbit Coupling Parameter}
Both standard Hall effect measurements \cite{Hatke2017,Thomas2018,Tschirky2017,Kim2010,Lee2011} and a magnetic field modulation technique \cite{Holmes2016} were employed to quantify the spin-orbit coupling in the 16 nm InAs QW (Sample B). The magnetoresistance oscillations of $\rho_{xx}$ in response to the applied field are clearly visible in Fig. \ref{fig:rashba}(a). However, when the FFT is taken of the data, as depicted in black in Fig. \ref{fig:rashba}(b), only a single peak is evident. The lack of a split peak may be due to a weak Rashba response. To further investigate any potential Rashba splitting and to improve measurement sensitivity, the modulated magnetic field technique was utilized to directly measure $dR_{xx}/dB$ and $d^2R_{xx}/dB^2$. A secondary solenoid, collinear with the primary DC field-generating solenoid, provided a small AC field in addition to the DC field. The AC field utilized to measure $dR_{xx}/dB$ had a magnitude on the order of 1.6 mT and a frequency of 33 Hz. The resulting $dR_{xx}/dB$ signal is depicted versus the applied static field in light blue in Fig. \ref{fig:rashba}(a) \cite{Holmes2016}. The top gate voltage was kept at $V_{g}=0$ V and the sample carrier density was measured to be $3.38\times10^{11}\ /$cm$^2$, corresponding to a mobility of $9.70\times10^5$ cm$^2$/Vs. A clear beating pattern is present in the low field oscillation data. The FFT of the low field oscillations is depicted in Fig. \ref{fig:rashba}(b) in light blue. Two peaks appear, nearly equal in frequency, arising from spin splitting \cite{Kim2010,Lee2011,Datta1990,Gauer1996}.

To quantify the peak splitting, we switch to units of inverse field where the oscillations are periodic. Next, an FFT is taken of the oscillations. The result of the Fourier transform is displayed in Fig. \ref{fig:rashba}(b). The two spin-split channels, labeled $n_-$ and $n_+$ are highlighted by the vertical dashed lines. The frequency corresponding to the peak of each channel is extracted and can be converted to the carrier density utilizing the Onsager relation to get $n_i=qf_i/h$ \cite{Onsager1952}, where $i=+,-$. We can extract the total carrier density, $n_T=n_++n_-=3.2\times10^{11}\ /$cm$^2$, giving good agreement with the carrier density measured by Hall transport. From the density of $n_+$ and $n_-$ we can now extract the total SOC parameter, $\Omega$, using the relationship \cite{Engels1997,Datta1990}:
\begin{equation}
    \Omega=\frac{\Delta n \hbar^2}{m^*}\sqrt{\frac{\pi}{2(n_T-\Delta n)}}
    \label{eq:alpha}
\end{equation}
Where $\Delta n=n_+-n_-$. Using Eq. \ref{eq:alpha} we find $\Omega=12.0\times10^{-12}$ eV m. This value is similar to the Rashba parameters found for InAs QWs in literature \cite{Thomas2018,Hatke2017,Grundler2000}, suggesting that the inherent asymmetry in the structure provides a small built-in electric field to the quantum well. Additionally, we can extract the spin-orbit length using $l_{SO}=\frac{1}{\Delta n}\sqrt{(n_T-\Delta n)/2\pi}$ \cite{Dettwiler2017,Hatke2017}. We find a spin-orbit length of $l_{SO}=102$ nm. This value is similar to the $l_{SO}$ values found by Hatke \textit{et al.} despite the lack of an applied electric field \cite{Hatke2017}.

\section{\label{sec:conc}Conclusion}
In conclusion, we have demonstrated the advantages of utilizing strain compensation in InAs QWs grown on mismatched InP substrates. The addition of tensile strain in the InGaAs cladding layer increases the critical thickness of the InAs well, enabling $t_{QW}$ to approach the QW thickness of InAs grown on nearly lattice-matched GaSb substrates. By increasing the thickness of the QW we demonstrated, utilizing 1D Schr\"odinger-Poisson simulations, that the percentage of the 2DEG state contained within the InAs layer increases from 50$\%$ up to $97\%$ for our thickest InAs QW samples.

Magnetotransport measurements indicated that the lower bound of the In concentration in our 10 nm thick InGaAs cladding layers is $x=0.72$. When the In concentration was lowered beyond $x=0.72$, the QW mobility dropped for all thicknesses of the QW. Our In$_{0.72}$Ga$_{0.28}$As QW with $t_{QW}=16$ nm showed the highest mobility with a record-breaking carrier mobility of $\mu=1.16\times10^6$ cm$^2/$Vs with $n_s=4.2\times10^{11}$ /cm$^2$. Next, we used the temperature dependence of the SdH oscillations to extract the effective mass and quantum lifetime, finding: $m^*=0.031m_e$ and $\tau_q=0.45$ ps, which is consistent with previously measured values. Finally, utilizing a modulated field measurement we were able to extract the SOC parameter, $\Omega=12.0\times10^{-12}$ eV m and the spin-orbit length, $l_{SO}=102$ nm. This work demonstrates the potential of intentionally incorporating strain into the QW cladding layers to maximize the critical thickness of the QW region for  InAs QWs grown on mismatched InP substrates.

\section{\label{sec:Ack}Acknowledgements}
The authors acknowledge financial support of the growth and characterization of these quantum wells by the US Department of Energy under award No. DE-SC0019274 and the UK Science and Technology Facilities Council under award No. ST/Y005074/1. We acknowledge the use of the shared facilities of the NSF Materials Research Science and Engineering Center (MRSEC) at the University of California Santa Barbara (Grant No. DMR 2308708) and the fabrication facilities of the UCSB Nanofabrication Facility, an open access laboratory.


\bibliography{apssamp}

\end{document}